\pdfoutput=1
\documentclass[aps,preprint,longbibliography,superscriptaddress]{revtex4-1}  

\usepackage{graphicx}  
\usepackage{dcolumn}   
\usepackage{bm}        
\usepackage{amssymb}   
\usepackage{amsmath}
\usepackage{bm}
\usepackage{afterpage}

\usepackage[caption=false,subrefformat=parens,labelformat=parens]{subfig}
\begin{document}


\title{Wrinkling instability of an inhomogeneously stretched viscous sheet}


\author{Siddarth Srinivasan}
\author{Zhiyan Wei}
\affiliation{John A. Paulson School of Engineering and Applied Sciences, Harvard University,  Cambridge, MA 02138, USA}
\author{L. Mahadevan}
\email[Corresponding author:]{lmahadev@g.harvard.edu}
\affiliation{John A. Paulson School of Engineering and Applied Sciences, Harvard University,  Cambridge, MA 02138, USA}
\affiliation{Departments of Physics, and Organismic and Evolutionary Biology, Kavli Institute and  Wyss Institute, Harvard University, Cambridge, Massachusetts 02138, USA}



\date{\today}

\begin{abstract}
Motivated by the redrawing of hot glass into thin sheets, we investigate the shape and stability of a thin viscous sheet that is inhomogeneously stretched in an imposed non-uniform temperature field. We first determine the associated base flow by solving the long-timescale stretching flow of a flat sheet as a function of two dimensionless parameters: the normalized stretching velocity $\alpha$, and a dimensionless width of the heating zone $\beta$. This allows us to determine the conditions for the onset of an out-of-plane wrinkling instability stated in terms of an eigenvalue problem for a linear partial differential equation governing the displacement of the midsurface of the sheet. We show the sheet can become unstable in two regions that are upstream and downstream of the heating zone where the minimum in-plane stress is negative. This yields the shape and growth rates of the most unstable buckling mode in both regions for various values of the stretching velocity and heating zone width. A transition from stationary to oscillatory unstable modes is found in the upstream region with increasing $\beta$ while the downstream region is always stationary. We show that the wrinkling instability can be entirely suppressed when the surface tension is large enough relative to the magnitude of the in-plane stress. Finally, we present an operating diagram that indicates regions of the parameter space that result in a required outlet sheet thickness upon stretching, while simultaneously minimizing or suppressing the out-of-plane buckling; a result that is relevant for the glass redraw method used to create ultrathin glass sheets.

\end{abstract}


\maketitle


\section{Introduction}
The flow of thin viscous fluid sheets has been widely studied in the context of various industrial and geophysical processes \cite{FLM:8496051,PhysRevE.68.036305}. In these flows, the interplay between bending and stretching of the thin fluid sheet often gives rise to folding and buckling instabilities that are observed in phenomena that span orders of magnitude in length. Examples range from geophysical processes such as the organization of supraglacial lakes \cite{GRL:GRL28770}, the buckling of layered geological strata \cite{johnson1994folding}, deformation of the lithosphere and subduction zones, \cite{England01081982,TECT:TECT2207} and surface folding of p\={a}hoehoe lava flows \cite{FINK1978151}, to more mundane everyday phenomena such as the folding of a sheet of honey \cite{skorobogatiy2000} and the wrinkling of the skin of scalded milk \cite{B516741H}. Thin fluid sheet flows are also relevant in industrial applications involving shaping, moulding, extrusion, film casting and film blowing processes, and are particularly important in the manufacture of flat glass by the float-glass processes \cite{pilkington1969}, the overflow downdraw or fusion processes \cite{m1967sheet}, and the redraw processes \cite{kiely_jfm_2015}. In this last technique, pre-cast sheets of molten glass are simultaneously heated in a furnace and drawn via a tensile force to obtain ultra-thin glass sheets with typical thicknesses that are $<100 ~\mu$m, a necessity for many modern haptic technologies.  However, an injudicious choice of stretching rates or applied heating profiles can give rise to wrinkles that, given the small thickness, adversely affect the uniformity of the sheet. An example of these wrinkles in a glass sheet with varying thickness that results from the redraw process is shown in Fig. \ref{fig:schematic}. Consequently, understanding the formation, size and shape of these instabilities, and determining the set of process parameters that suppresses the instability is of great practical importance in achieving ultra-thin flat glass sheets, and is the primary motivation of this paper.

Models for the dynamics of thin viscous sheets have focused on reduced order `viscous plate' theories, where the full incompressible Navier-Stokes equations are asymptotically reduced to equations that govern the bending and stretching of the center-line for thin sheets \cite{FLM:373441,howell1994extensional,EJM:2321580,FLM:101139}. Viscous plate models have been primarily used in studying buckling, coiling and folding phenomena in sheets with uniform viscosities. The exceptions include the work of Pfingstag, Audoly \& Boudaoud \cite{pfingstagpof2011} who have studied two specific two-dimensional examples involving non-homogeneous viscosity in thin sheets: necking induced by in-plane viscosity variations, and the out-of-plane deformation under imposed transverse variations in viscosity. Filippov \& Zheng \cite{filippov2010} determined the boundary shape and thickness distribution of three-dimensional non isoviscous sheets under a stretching flow in the redraw process, for two model temperature profiles, and showed the existence of unstable compressive zones. However, they do not solve for the out-of-plane deformation to determine the shape of the unstable modes. Also related is the work of Perdigou \& Audoly \cite{Perdigou2016291}, who investigate the problem of a falling viscous sheet under the action of gravity, and determine the stability and out-of-plane modes for a constant thickness and viscosity.

In this paper, motivated by the glass redraw process, we investigate the shape and stability of three-dimensional thin non-homogeneous viscous sheets in the redraw processes. The aim of this work is to understand how the region of instability in the sheet varies with the draw rate and the size and shape of the heating zone, and to compute the resulting shape of the most unstable modes in these zones. We approach the problem in two parts. First, we determine the steady-state shape and plane deformation of an initially flat sheet, analogous to the work of Filippov \& Zheng \cite{filippov2010}, for a range of different temperature and outlet velocities. Then, using this flat deformed sheet as the base-state solution, we formulate and solve a linearized eigenvalue problem to determine the out-of-plane deformation of the mid-plane, and show that the eigenmodes correspond to a viscous buckling instability. In Section \ref{section2}, we introduce the geometry and variables of interest, and state the equations governing both the in-plane steady-state flow, and the out-of-plane buckling of the center-line. In Section \ref{section3}, we determine the base-state solutions as well as the out-of-plane deformations for the most unstable buckling mode for different values of the operating parameters. We discuss the effect of surface tension and determine the region of parameter space that results in a specified outlet sheet thickness while either minimizing or eliminating the out-of-plane deformation. Finally, we provide a summary of our results and discuss its implications in the manufacture of ultra-thin glass via the redraw technique.





\section{Geometry and Variables}
\label{section2}

\begin{center}
\begin{figure}[t]
\includegraphics[width=\columnwidth]{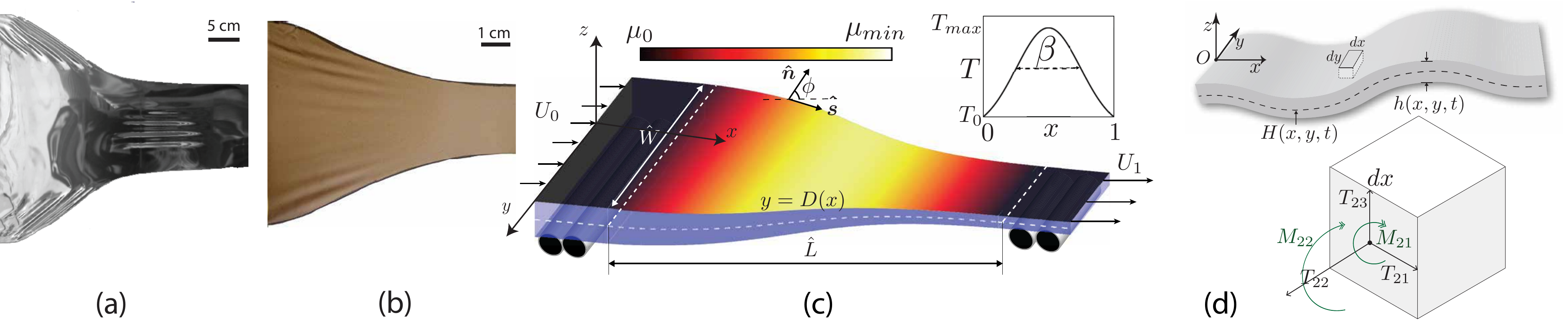}
\caption{\label{fig:schematic} (a) Example of a heated sheet of thin glass that is undergoing the redraw process. Wrinkles are observed parallel to the edges and in center of the sheet. Note the tensile wrinkles in the downstream end of the glass, possibly due to strong inhomogeneity induced by temperature. (b) An analogous experiment in an elastic setting, carried out by stretching a thin nitrile sheet of a uniform thickness of $0.10$ mm cut into the shape of the glass sheet with a longitudinal strain applied to the clamped edge shows elastic wrinkles; these vanish when the boundary stresses are relaxed, unlike in the case of the glass sheet. (c) Schematic of the re-draw process for a thin viscous liquid with the inlet feed velocity $\hat{U}_0$ and the outlet draw velocity $\hat{U}_1$. Here, $y=D(x)$ is the lateral free boundary. A gaussian heating profile is prescribed in the furnace zone (see inset). The colormap indicates the variation of the viscosity of the sheet. (d) Top: Illustration of the out-of-plane deformation of the sheet where the thickness is $h(x,y,t)$ and the mid-surface displacement is $H(x,y,t)$. Bottom: The components of the resultant membrane stresses $\boldsymbol{T}$ and twisting and bending moments $\boldsymbol{M}$ on a unit volume.}
\end{figure}
\end{center}

We consider small deformations of nearly flat thin sheets undergoing moderate out-of-plane rotations in the redraw process, with a configuration as shown in Fig.~\ref{fig:schematic}(c). Hat superscripts indicate dimensional quantities and plain variables denote dimensionless quantities. We use a coordinate system where $\hat{x}$ denotes the position directed along the length axis, $\hat{y}$ along the width axis and $\hat{z}$ along the height axis. The origin is located on the undeformed centerline at the inlet and is indicated by the point $O$ in Fig.~\ref{fig:schematic}(d). The total length of the redraw zone is $\hat{L}$, the initial width of the sheet is $\hat{W}$ and the initial thickness is $\hat{h}_0$. The slenderness ratio is defined as $\epsilon=\hat{h}_0/\hat{L}\ll 1$. The sheet enters the inlet at $\hat{x}=0$ with a velocity $\hat{U}_0$ and is initially at a temperature $\hat{T}_0$ before it flows to the heating zone of the furnace where the viscosity decreases as illustrated in Fig.~\ref{fig:schematic}(c).  In this study, we ignore the effects of fluid inertia and consider the scenario of small Reynolds number $Re=\hat{\rho}\hat{U}_0\hat{L}/\hat{\mu}_0 \ll 1$, where $\hat{\rho}$ is the density and $\hat{\gamma}$ is the surface tension of the fluid that are assumed to be constant. $\hat{\mu}_0$ is a characteristic viscosity of the fluid defined in Section \ref{sec:temp_profiles}. Typical values of these parameters in the redraw process are $\hat{\rho} = 10^3$ kg/m$^3$, $\hat{U}_0 = 10^{-4}$ m/s, $\hat{L} = 1$ m and $\hat{\mu}_0 = 10^6 $ Pa.s, leading to $Re \sim 10^{-7}$. We work in the regime of small Stokes number $St=(\hat{\rho}_0 g \hat{L}^2)/(\hat{\mu}_0 \hat{U}_0) \ll 1$, so that we neglect the effect of gravity $g$ (that is acting along the $\hat{x}$-axis) relative to the viscous shear.

As discussed by Howell \cite{howell1994extensional}, there exist two different classes of models describing the deformation of thin viscous sheets. These models are obtained by applying scaling assumptions valid at different time scales. Over long time-scales $\hat{t}\sim \hat{L}/\hat{U}$, the Trouton model is used to describe the planar flow where the mid-surface remains two-dimensional and the sheet thickness and shape deforms purely by stretching. At short time scales $\hat{t}\sim \epsilon^2 \hat{L}/\hat{U}$, the Buckmaster-Nachman-Ting (BNT) scaling \cite{FLM:373441} describes the incipient out-of-plane deformation of the mid-surface of the thin sheet. Before describing the dimensionally reduced thin plate equations that are asymptotically valid in the slenderness parameter $\epsilon=\hat{h_0}/\hat{L}$, we first scale the in-plane coordinates as $\hat{x}=x\hat{L}$, $\hat{y}=y\hat{L}$ and in-plane velocities as $\hat{u}=u \hat{U}_0$, $\hat{v}=v \hat{U}_0$. The out-of-plane coordinate $\hat{z}$, the thickness $\hat{h}$ and the center-line location $\hat{H}$ are scaled with the slenderness ratio $\epsilon$ as $\hat{z}=z \epsilon \hat{L}$, $\hat{h}=h \epsilon \hat{L}$ and $\hat{H}=H\epsilon \hat{L}$, where $\epsilon \ll 1$. 

In the heating zone, an external temperature profile $\hat{T}(\hat{x})$ is applied (see the inset of Fig.~\ref{fig:schematic}(c)). We assume that the fluid is in radiative equilibrium with the furnace so that the temperature in the viscous sheet is identical to that prescribed by the external heating device. This results in a prescribed non-uniform viscosity field $\hat{\mu}(\hat{x})$ across the length of the fluid sheet. At the outlet, which is located at $\hat{x}=\hat{L}$, the sheet is drawn at a large constant velocity $\hat{U}_1$ and cooled to the initial temperature $\hat{T}_0$. The stretching action of the draw roller at the oulet gives rise to a velocity field $\vec{U}=(\hat{u},\hat{v},\hat{w})$, where $\hat{u}, \hat{v}$ and $\hat{w}$ are the spatially varying velocity components along the $\hat{x}$, $\hat{y}$ and $\hat{z}$ directions respectively. Consequently, tensile and compressive stresses generated by the flow will result in a thinner and laterally contracted viscous sheet downstream of the heating zone as illustrated in Fig.~\ref{fig:schematic}(c). The dimensionless stretching velocity is defined as

\begin{equation}
\label{eq:alpha_def}
\alpha=\hat{U}_1/\hat{U}_0
\end{equation}

where $\alpha > 1$ is a measure of the magnitude of applied extensional flow in the redraw process. The deformation of the thin sheet is driven by the viscous stretching and bending stresses that are generated during the flow and are depend on the values of $\alpha$ and choice of $\hat{T}(\hat{x})$. The ensuing dynamics of sheet deformation are described by the evolution of the sheet thickness $\hat{h}(\hat{x},\hat{y},\hat{t})$, and the location of the mid-surface $\hat{H}(\hat{x},\hat{y},\hat{t})$. During deformation, the material of the sheet is confined between the surfaces $\hat{z}_\pm= \hat{H} \pm \hat{h}/2$. Our goal is to understand how the dimensionless stretching velocity $\alpha$ and applied heating profile $\hat{T}(\hat{x})$ can be chosen to obtain a desired steady state thickness distribution $\hat{h}(x,y)$ and a steady planar two-dimensional flow field $\hat{u}(\hat{x},\hat{y})$, $\hat{v}(\hat{x},\hat{y})$ in the redraw process, that are also stable to any out-of-plane deformations that arise from perturbations of the mid-surface $\hat{H}(\hat{x},\hat{y})$ and therefore result in flat thin sheets.  \\

\subsection{Base State}

In the absence of out-of-plane deformations, the flow remains nearly two-dimensional and planar and the mid-surface is constant, \textit{i.e.} $H(x,y)=0$. The three unknown variables that describe the steady base state are the thickness field $h(x,y)$ and in-plane velocity fields $u(x,y)$ and $v(x,y)$. In this planar state, the in-plane velocities are independent of the z-coordinate. Therefore, we require three governing equations to fully determine the steady state values of $h$, $u$ and $v$. The first relation is provided by the conservation of volume for an incompressible liquid at steady state and is given by, 

\begin{equation}
\label{eq:continuity}
(uh)_x+(vh)_y=0
\end{equation}

where, $(.)_x=\partial (.)/\partial x$, etc. Then, for purely stretching flows of an initially flat sheet, a balance of the Cauchy stress $\boldsymbol{\sigma}$ provides the remaining two governing equations that are written in terms of the resultant membrane stress $\boldsymbol{T}=\int_{z^-}^{z^+} \boldsymbol{\sigma} dz$ at steady state \cite{howell1994extensional},

\begin{gather}
\label{eq:trouton_dimless_a}
(T_{11})_x+(T_{12})_y=0 \\
\label{eq:trouton_dimless_b}
(T_{12})_x+(T_{22})_y=0.
\end{gather}

where, $T_{11}=2 \mu h(2u_x +v_y)$, $T_{12}=\mu h(u_y +v_x)$ and $T_{22}=2 \mu h(u_x +2v_y)$ are the components of the membrane stress $\mathbf{T}$ for a flat sheet at steady state, with $H=0$. The boundary conditions for the base-state are given by the fixed velocities at the inlet and outlet, and the no-flux and stress-free boundary condition at the lateral edge,

  \begin{equation}
    \begin{aligned}
\label{eq:trouton_bc_a}
u=1;~v=0 \quad &\text{at } x=0\\
u=\alpha;~v=0 \quad &\text{at } x=1\\
\boldsymbol{\hat{n}}\cdot\boldsymbol{u}=0;~\boldsymbol{\hat{n}} \cdot \boldsymbol{T}=0 \quad &\text{at } y=D(x).
\end{aligned}
\end{equation}

where, the sheet is initially at a uniform thickness of $h=1$, $\alpha$ is the draw ratio defined in (\ref{eq:alpha_def}), $y=D(x)$ is the free lateral edge of the sheet at steady state, $\boldsymbol{\hat{n}}$ is the outer normal vector at the free lateral edge (see Fig.~\ref{fig:schematic}(c)), and $\mathbf{u}=(u,v)$ is the velocity vector. 

\subsection{Out-of-plane deformation}
 
We state the general reduced order thin plate equations governing the deformation of nearly flat thin sheets in the absence of inertia, and direct the reader to the work of Howell \cite{howell1994extensional}, Slim \textit{et al}. \cite{slim_2012_jfm}, and the recent work of Pfingstag, Audoly \& Boudaoud \cite{pfingstag_2011_jfm} for detailed asymptotic derivations. The four unknown fields during out-of-plane deformations are the mid-surface displacement $H(x,y,t)$, the thickness $h(x,y,t)$, and the mean in-plane velocities $\bar{u}(x,y)$ and $\bar{v}(x,y)$. Here, the mean velocities are defined as $\bar{u}=1/h \int_{z^-}^{z^+} u dz$ and $\bar{v}=1/h \int_{z^-}^{z^+} v dz$ where $u$ and $v$ now depend on the z-coordinate in contrast to the base state. In the BNT scaling \cite{FLM:373441}, the equation governing the conversation of volume reduces to,

\begin{equation}
h_t=0
\end{equation}

The leading order thin plate governing equations are expressed as in terms of the resultant membrane stresses $\boldsymbol{T}=\int_{z^-}^{z^+} \boldsymbol{\sigma} dz $ and the bending and twisting moments $\boldsymbol{M}=\int_{z^-}^{z^+} (z-H) \boldsymbol{\sigma} dz$ as,

\begin{gather}
\label{eq:BNT_dimless_a}
(T_{11})_x+(T_{12})_y=0 \\
\label{eq:BNT_dimless_b}
(T_{12})_x+(T_{22})_y=0 \\
\label{eq:BNT_dimless_c}
(M_{11})_{xx}+2(M_{12})_{xy}+(M_{22})_{yy}+H_{xx}T_{11}+2H_{xy}T_{12}+H_{yy}T_{22}+\Gamma(H_{xx}+H_{yy})=0.
\end{gather}

\newsavebox\Telastic
\begin{lrbox}{\Telastic}
  \begin{minipage}{0.4\textwidth}
  
  \begin{align*} 
T_{11} &=2Gh \left(2\bar{u}_{x}+ \bar{v}_{y}+H^2_x + \frac{1}{2}H^2_y\right)\\
T_{12} &=G h\left(\bar{u}_{y}+ \bar{v}_{x}+H_x H_y\right)\\
T_{22} &=2Gh \left(\bar{u}_{x}+ 2 \bar{v}_{y}+\frac{1}{2}H^2_x + H^2_y\right) 
\end{align*} 
  \end{minipage}
\end{lrbox}

\newsavebox\Tviscous
\begin{lrbox}{\Tviscous}
  \begin{minipage}{0.4\textwidth}
  \begin{align}
T_{11} &=2\mu h(2\bar{u}_x+\bar{v}_y+2H_{tx}H_x+H_{ty}H_y)\label{eq:T11}\\
T_{12} &=\mu h(\bar{u}_y+\bar{v}_x+H_{tx}H_y+H_{ty}H_x) \label{eq:T12}\\
T_{22} &=2\mu h(\bar{u}_x+2\bar{v}_y+H_{tx}H_x+2 H_{ty}H_y)\label{eq:T22}
\end{align}
  \end{minipage}
\end{lrbox}

\newsavebox\Melastic
\begin{lrbox}{\Melastic}
  \begin{minipage}{0.4\textwidth}
  \begin{align*} 
M_{11} &=-\frac{G h^3}{6}\left(2 H_{xx}+ H_{yy}\right)\\
M_{12} &=-\frac{G h^3}{6}H_{xy}\\
M_{22} &=-\frac{G h^3}{6}\left(H_{xx}+ 2 H_{yy}\right)
\end{align*}
\end{minipage}
\end{lrbox}

\newsavebox\Mviscous
\begin{lrbox}{\Mviscous}
  \begin{minipage}{0.4\textwidth}
  \begin{align} 
M_{11} &=-\frac{\mu h^3}{6}\left(2H_{xxt}+H_{yyt} \right)\label{eq:M11}\\
M_{12} &=-\frac{\mu h^3}{6}H_{xyt} \label{eq:M12}\\
M_{22} &=-\frac{\mu h^3}{6}\left(H_{xxt}+2H_{yyt} \label{eq:M22} \right)
\end{align}
\end{minipage}
\end{lrbox}

\begin{table}
\caption{\label{tab:table1}Relations for the resultant membrane stresses $\boldsymbol{T}$ and the bending and twisting moments $\boldsymbol{M}$ for thin elastic sheets with a poisson ratio of $\nu=1/2$ and thin viscous sheets, where $H(x,y)$ is the location of the midsurface and $h(x,y)$ is the thickness of the sheet, as shown in Fig.~(\ref{fig:schematic}). For elastic sheets, $\bar{u}$ and $\bar{v}$ are the mean in-plane displacements, $G$ is the shear modulus. For viscous sheets, $\bar{u}$ and $\bar{v}$ are the mean z-integrated in-plane velocities, where $\bar{u}=1/h \int_{H-h/2}^{H+h/2} u dz$ and $\bar{v}=1/h \int_{z^-}^{z^+} u dz$, and $\mu$ is the shear viscosity.}
\begin{ruledtabular}
\resizebox{\columnwidth}{!}{
\begin{tabular}{l@{}c@{}c@{}}
 &\multicolumn{1}{c}{Elastic}&\multicolumn{1}{c}{Viscous)
}\\\\[0.1cm]

$\underline{\boldsymbol{T}}$ & \usebox{\Telastic} & \usebox{\Tviscous}\\[2cm]

$\underline{\boldsymbol{M}}$ & \usebox{\Melastic} & \usebox{\Mviscous} \\
\end{tabular}
}
\end{ruledtabular}
\end{table}

where $T_{11}, T_{12}, T_{22}$ and $M_{11}, M_{12}, M_{22}$ are the components of resultant stresses and moments (see inset of Fig~\ref{fig:schematic}d), and in contrast to the base state, now include non-linear terms due to finite rotations of the mid-surface, \textit{i.e.} $H \neq 0$. Equations (\ref{eq:BNT_dimless_a})-(\ref{eq:BNT_dimless_c}) are independent of the constitutive law for the thin sheets and arise solely from force equilibrium and dimensional scaling considerations. (\ref{eq:BNT_dimless_a}) and (\ref{eq:BNT_dimless_b}) result from a balance of stresses in the in-plane directions and (\ref{eq:BNT_dimless_c}) balances the bending stress due to out-of-plane displacements with contribution of the in-plane stresses that can either drive or stabilize out-of-plane deformations. The last term in (\ref{eq:BNT_dimless_c}) is due to the contribution of surface tension $\hat{\gamma}$ in stabilizing out-of-plane displacement of the sheet, where $\Gamma=2\hat{\gamma}/(\hat{\mu}_0 \hat{U}_0)$ is an inverse capillary number, and where $\hat{\mu}_0$ is the reference viscosity whose value will be defined later in Section \ref{sec:temp_profiles}. To emphasize the general applicability of these equations to thin sheet geometries, we have provided expressions that relate $\boldsymbol{T}$ and $\boldsymbol{M}$ to the kinematic variables in Table~\ref{tab:table1} for two constitutive laws: an isotropic linear elastic solid and a Newtonian liquid. In the former case, (\ref{eq:BNT_dimless_a})-(\ref{eq:BNT_dimless_c}) reduce to the F\"{o}ppl von-K\'{a}rm\'{a}n equations for thin elastic sheets. For Newtonian fluids, they represent the viscous sheet model \cite{howell1994extensional,slim_2012_jfm,pfingstag_2011_jfm} that is obtained using the scaling introduced by Buckmaster, Nachman and Ting (BNT) \cite{FLM:373441}. The close similarity in the expressions for the resultant membrane stresses and moments in Table.~\ref{tab:table1} is a manifestation of the Stokes-Rayleigh analogy between isotropic elastic solid with a Poisson ratio $\nu=1/2$ and the zero Reynolds number fluid flow. 


To complete the formulation of the boundary value problem in (\ref{eq:BNT_dimless_a})-(\ref{eq:BNT_dimless_c}), we need boundary conditions at the inlet and the outlet. At $x=0$ and $x=1$, we use the clamped boundary conditions,

\begin{equation}
\label{eq:eigenprob_bc_a}
H=0 \quad \text{and} \quad H_x=0.
\end{equation}

At the free edges of the sheet, we consider a local orthogonal axis along the exterior normal $\boldsymbol{\hat{n}}$ and tangential direction $\boldsymbol{\hat{s}}$ (Fig.~\ref{fig:schematic}c). We obtain the natural boundary conditions by considering the boundary virtual work line integral along a segment C of the boundary \cite{timoshenko1959theory},

\begin{equation}
\label{eq:work_conjugate}
W=\int\limits_C \left(T_n H + M_{ns} \frac{\partial H}{\partial s}+ M_{nn} \frac{\partial H}{\partial n}\right) ds + \int\limits_C \left(\Gamma_n H\right) ds
\end{equation}

The first term corresponds to the virtual work due to internal transverse stresses and bending moments. The last term arises from the virtual work arising from external capillary forces. Here, $\partial/\partial n$ and $\partial/\partial s$ indicate derivatives along the normal and tangent direction, and $M_{nn}$ and $M_{ns}$ are the components of the moment at the lateral edge and can be expressed as,
\begin{align}
M_{nn}&=M_{11} \cos\phi^2 +2M_{12} \cos\phi \sin\phi + M_{22} \sin\phi^2 \\
M_{ns}&=(M_{22}-M_{11})\cos\phi\sin\phi+M_{12}(\cos^2\phi-\sin^2\phi).
\end{align}

where $\phi$ is the angle between the exterior normal vector $\boldsymbol{\hat{n}}$ and the x-axis and $\boldsymbol{\hat{s}}$ is the tangent vector as shown in Fig.~\ref{fig:schematic}(c). $T_n$ is the z-integrated transverse stress along the boundary plane,

\begin{equation}
T_n=T_{13}\cos\phi + T_{23} \sin \phi.
\end{equation}

where,
\begin{align}
\label{eq:mT13}
T_{13}&=(M_{11})_x +(M_{12})_y+H_x T_{11}+H_y T_{12} \\
T_{23}&=(M_{12})_x+(M_{22})_y +H_x T_{12}+H_y T_{22}.
\end{align}

and $\Gamma_n ds$ is the projection of the curvature force associated with surface tension in the transverse direction along the boundary plane.
\begin{equation}
\Gamma_n= \Gamma \left(H_x  \cos \phi + H_y \sin \phi\right).
\end{equation}
 
The natural boundary conditions are then obtained by setting the work-conjugate variables obtained upon integration by parts of (\ref{eq:work_conjugate}) to 0. As we consider only one-half of the sheet in our simulations, the symmetry boundary conditions at $y=0$ are then,
\begin{equation}
\label{eq:eigenprob_bc_c}
H_y=0 \quad \text{and} \quad T_n-\frac{\partial M_{ns}}{\partial s} + \Gamma_n=0.
\end{equation}

The free edge boundary condition at the lateral edge $y=D(x)$ where both the modified transverse shear stress and bending moments vanish are expressed as,
\begin{equation}
\label{eq:eigenprob_bc_d}
M_{nn}=0 \quad \text{and} \quad T_n-\frac{\partial M_{ns}}{\partial s}+\Gamma_n=0.
\end{equation}

The thickness $h(x,y)$, mid-surface location $H(x,y)$ and the mean in-plane velocities $\bar{u}(x,y)$ and $\bar{v}(x,y)$ are unknown and are coupled via the non-linear partial differential equations  (\ref{eq:BNT_dimless_a})-(\ref{eq:BNT_dimless_c}) subject to the boundary conditions (\ref{eq:eigenprob_bc_a}),(\ref{eq:eigenprob_bc_c}) and (\ref{eq:eigenprob_bc_d}). 

\subsection{Form of imposed temperature profiles.}
\label{sec:temp_profiles}

To complete the formulation, we also need the temperature field, which evolves rapidly compared to the viscous field. This implies that the temperature field $\hat{T}(\hat{x})$ in the fluid is equilibrated, and is modeled as a Gaussian of the form 

\begin{equation}
\label{eq:temp_profile}
\hat{T}(\hat{x}) =\hat{T}_0+ \Delta\hat{T} \exp\left({-\frac{(\hat{x}-\hat{L}/2)^2}{2\hat{\beta}^2}}\right)
\end{equation}

where $\hat{T}_0$ is the inlet temperature, $\Delta \hat{T}$ is the maximum temperature rise in the furnace and $\hat{\beta}$ is a measure of the width of the heating profile. In our simulations, we parameterize the applied temperature field by defining the scaled heating zone width

\begin{equation}
\beta=\frac{\hat{\beta}}{\hat{L}}.
\end{equation}

Therefore, the heating zone gets smaller with decreasing values of $\beta$. This results in a prescribed non-uniform viscosity field $\hat{\mu}(x)$ across the length of the fluid sheet. Motivated by the application to the glass re-draw process, and under the assumption of equilibration, we use the Fulcher equation \cite{JACE:JACE339} to exponentially relate the viscosity to the temperature field as,

\begin{equation}
\label{eq:viscosity_field}
\log_{10}\hat{\mu}(\hat{x})=-A+\frac{B}{\hat{T}(\hat{x})-C}
\end{equation}

where $A$, $B$ and $C$ are empirical constants that depends on the composition of the glass \cite{JACE:JACE339}, and $\hat{T}(\hat{x})$ is the glass temperature in Kelvin and $\hat{\mu}$ is the viscosity (in Pa.s). We assume Fulcher constants of $A=4.5$, $B=7500$ K and $C=520$ K. We define the scaled viscosity as $\mu = \hat{\mu}/\hat{\mu}_0$, where $\hat{\mu}_0$ is the viscosity at an arbitrary reference temperature chosen as $\hat{T}=1233$ K.
\section{Analysis}

\label{section3}
The shape and stability of the sheet is influenced by the outlet draw velocity and the shape of the temperature profile \cite{filippov2010}. We systematically vary two dimensionless variables that parameterizes these effects: the dimensionless stretching velocity $\alpha=\hat{U}_1/\hat{U}_0$ and the dimensionless heating zone width $\beta=\hat{\beta}/\hat{L}$. In the analysis below, we assume an inlet temperature of the molten glass as $\hat{T}_0=1123$ K and a maximum temperature rise of $\Delta \hat{T}=100$ K. This corresponds to the dimensionless viscosity varying from a maximum of $\mu \approx 80$ at the inlet at $x=0$ to a minimum of $\mu \approx 1$ at $x=0.5$. The numerical solutions are implemented using the Comsol Multiphysics\textsuperscript{\textregistered} finite element solver package. The base state solutions are computed using a deformed mesh method, and the eigenvalue problem and boundary conditions are implemented using the general form PDE module.
\subsection{Base State}
\label{sec:basestate}
\begin{figure}[t]
\centering
\subfloat[][]{\includegraphics[height=4 cm]{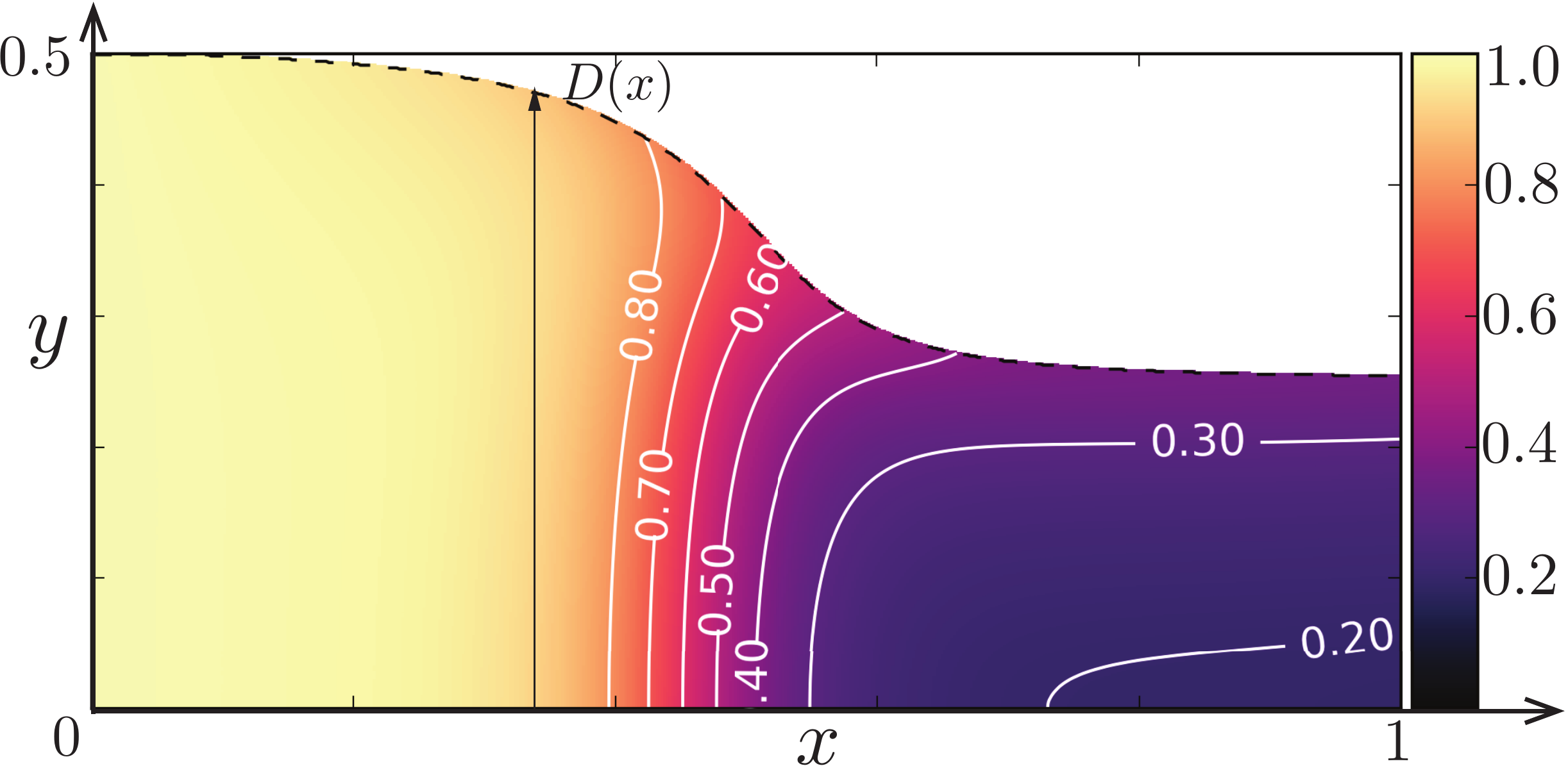}
\label{fig:heights}}
\subfloat[][]{\includegraphics[height=4 cm]{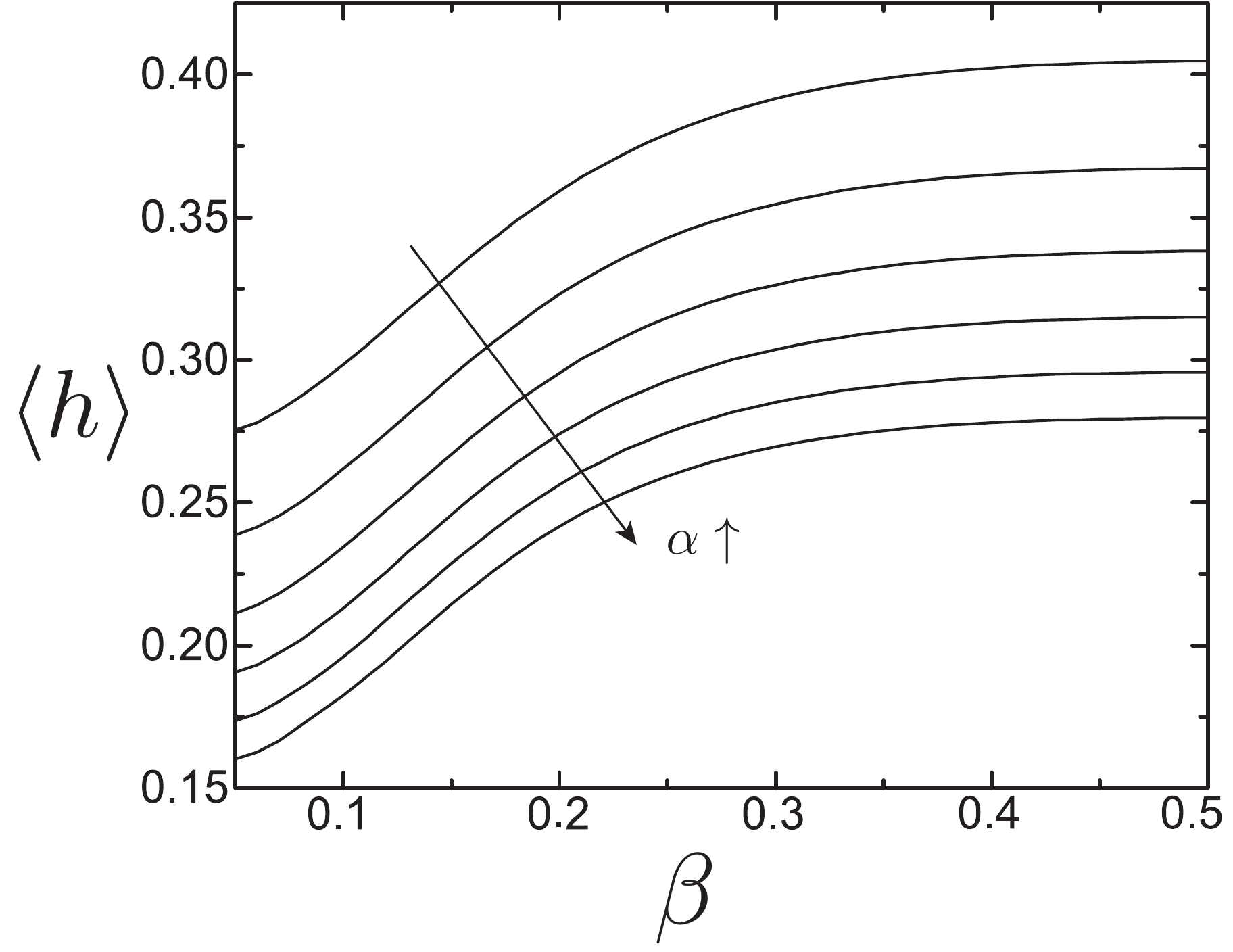}
\label{fig:height_map}}
\caption{\label{fig:combined_heights}Base state solutions of the flow. (a) Color plots indicating the shape and the normalized sheet thickness of a representative base state obtained by solving (\ref{eq:continuity})-(\ref{eq:trouton_dimless_b}) with $\alpha=8$ and $\beta=0.15$. The thickness $h(x,y)$ decreases from $h=1$ at $x=0$ for $\alpha>1$. The solid lines indicate contours of fixed thickness and the dotted line indicates the laterally contracted profile $D(x)$. (b) Plots of the variation of the averaged normalized sheet thickness $\langle h \rangle=1/D(1) \int_0^{D(1)} h dy$ evaluate at $x=1$ at the outlet with the the heating zone width $\beta$. Curves are shown for normalized stretching velocity $\alpha$ increasing in the direction of the arrow from 5 to 10 with increments in intervals of unity.}
\end{figure}

The governing equations (\ref{eq:continuity})-(\ref{eq:trouton_dimless_b}) for the base state are numerically solved to determine the steady state thickness field $h$ and the in-plane velocities $u$ and $v$ for each pair of values of $\alpha$ and $\beta$. In Fig.~\subref*{fig:heights}, we show the half profile of the sheet above the symmetry line $y=0$, and its thickness distribution for a representative value of $\alpha=8$ and $\beta=0.15$. The isocontours of  sheet thickness are marked by solid lines. The half-width profile of the sheet $D(x)$ is indicated by the dotted line. Thinning of the sheet and lateral contraction occurs near the heating zone in the region of low viscosity as expected. The thickness of the sheet is increased at the edges, consistent with the results of Filippov \& Zheng \cite{filippov2010}. To quantify the effective thickness as a result of the redraw process, we define the mean sheet thickness at $x=1$ as $\langle h \rangle=1/D(1) \int_0^{D(1)} h dy$. Here, $\langle h \rangle$ is determined for various values of $\alpha$ and $\beta$ and is shown in Fig.~\subref*{fig:height_map}. By increasing the stretching velocity or narrowing the heating zone, thinner sheets can be obtained. For example, at a fixed value of $\beta=0.10$, $\langle h \rangle$ decreases from 0.30 to 0.18 as $\alpha$ increases from 5 to 10. Decreasing the outlet sheet thickness by narrowing the width of the heating zone is most effective only for small values of $\beta$.\\

The stretching of the incompressible fluid sheet along its length induces contraction in the other two directions. Solving for the velocity profiles and the thickness allows us to determine the total in-plane stress $\boldsymbol{T}$ at every point in the sheet, where $\boldsymbol{T}$ consists of both normal and shear stress components. To visualize the state of tension and compression in the fluid sheet, we calculate the eigenvalues of $\boldsymbol{T}$ and denote the smaller eigenvalue by $T_1$ and the larger by $T_2$. A value of $T_1<0$ indicates that the sheet is locally under compression along the corresponding principal direction and under tension in the orthogonal principal direction. When both eigenvalues $T_1,T_2>0$, the sheet is under tension along both principal directions. For thin fluid sheets, compressive stresses drive buckling instabilities in the absence of stabilizing effects such as surface tension. Therefore, locations in the sheet where $T_1<0$ are prone to buckling induced by compressive stresses.\\

\begin{figure}[p!]
\centering
\includegraphics[width=0.8\columnwidth]{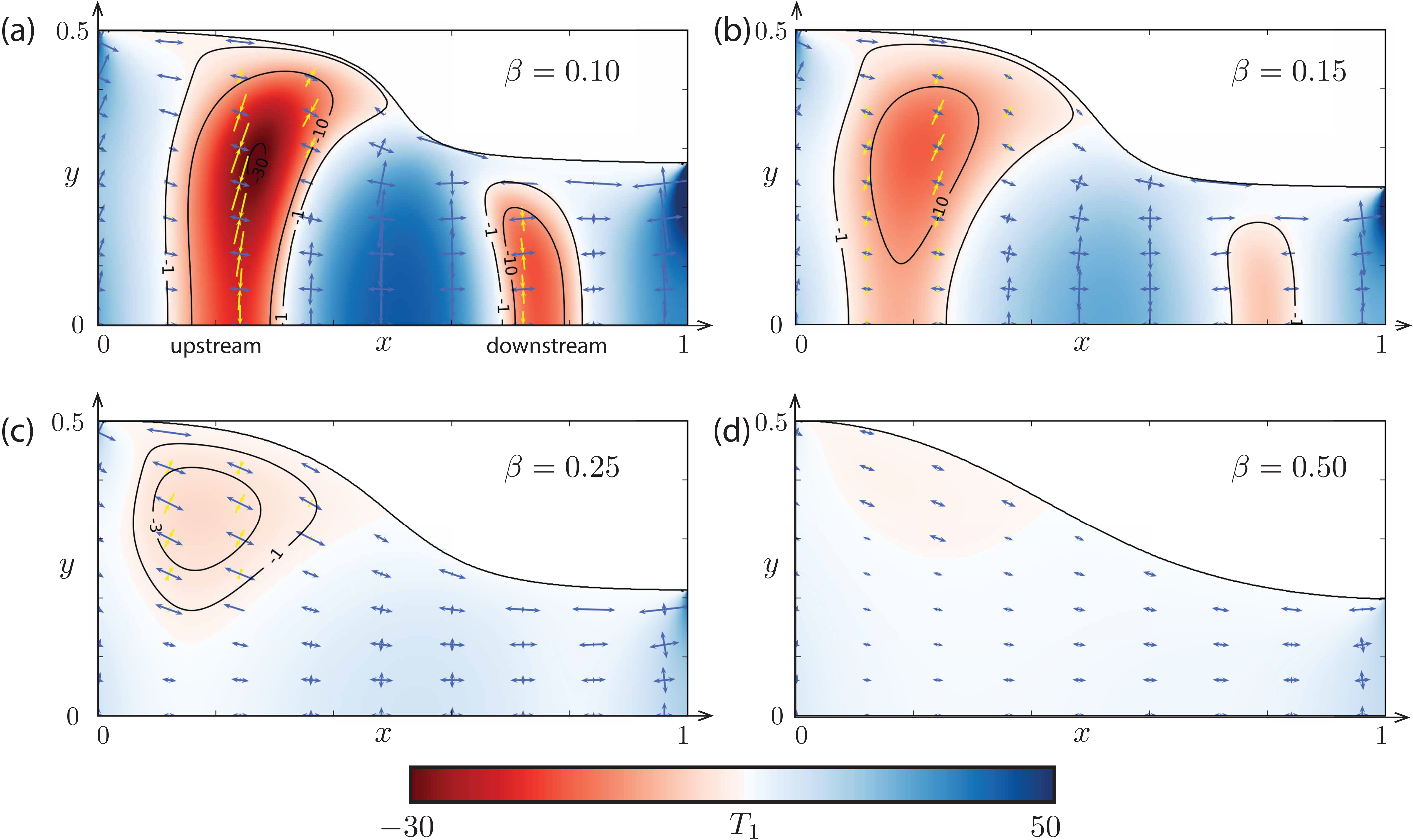}
\caption{\label{fig:stresses}Surface plots of the most negative eigenvalues $T_1$ of the local base-state stress tensor $\boldsymbol{T}$ for different heating zone widths $\beta$ at a fixed stretching velocity of $\alpha=8$ obtained by solving Eqs.~(\ref{eq:continuity})-(\ref{eq:trouton_dimless_b}). The plots in (a)-(d) are for values of $\beta=0.10$, 0.15, 0.20 and 0.50 respectively. The arrows correspond to the local eigenvectors of $\boldsymbol{T}$. The red regions indicates the presence of a compressive principal stress component ($T_1<0$) , while the blue regions indicate tensile stresses ($T_1>0,T_2>0$).}
\end{figure}

In Fig.~\ref{fig:stresses}(a)-(d), we show the effect of increasing the width of the heating zone on the position and size of the unstable zones. The surface plots show the magnitude of $T_1$ for values of $\beta=0.10,0.15,0.25$ and 0.50 respectively. The redraw velocity was fixed at $\alpha=8$. Similar results are obtained for other values of $\alpha$. The local principal directions of the stress tensor are indicated by the arrows with the length of the arrow proportional to the magnitude of the stress eigenvalue along the corresponding direction. Regions of the sheet that are under compression along one direction (i.e, $T_1<0$ \& $T_2>0$) are shown in red. In these regions the direction of compression is marked by yellow arrows. Regions where the sheet is under tension in both directions (i.e, $T_1>0$ \& $T_2>0$) are shown in blue. For $\beta=0.10$ and $0.15$, there are two unstable compressive zones, one located upstream near the inlet, and the other located downstream (\textit{i.e.} $x>0.5$) near the exit. Increasing the width of the heating zone has two main effects: (i) the size of both unstable compressive zones become smaller, and (ii) the downstream compressive zone entirely vanishes beyond a critical value of $\beta$. In our simulations, we observe that the sheet is always under tension in the region $x>0.5$ when $\beta > 0.2$ for $5 \leq \alpha \leq 10$, the range of draw ratios investigated. Finally, the magnitude of the maximum compressive stress diminishes in both regions when using a wider heating zone. As most of the sheet thinning and and lateral contraction occurs in the vicinity of the heating zone (see Fig.~\subref*{fig:heights}), the gradients in the in-plane velocities are localized to this region. A wider heating zone results in a larger region over which the velocity gradients manifest, leading to the reduced magnitude of the in-plane stresses. In the absence of surface tension, in thin viscous sheets, we qualitatively expect buckling instabilities to occur in the compressive zones with the wrinkles oriented perpendicular to the direction of compression.\\

\afterpage{\clearpage}

\begin{figure}[p!]
\centering
\includegraphics[width=0.8\columnwidth]{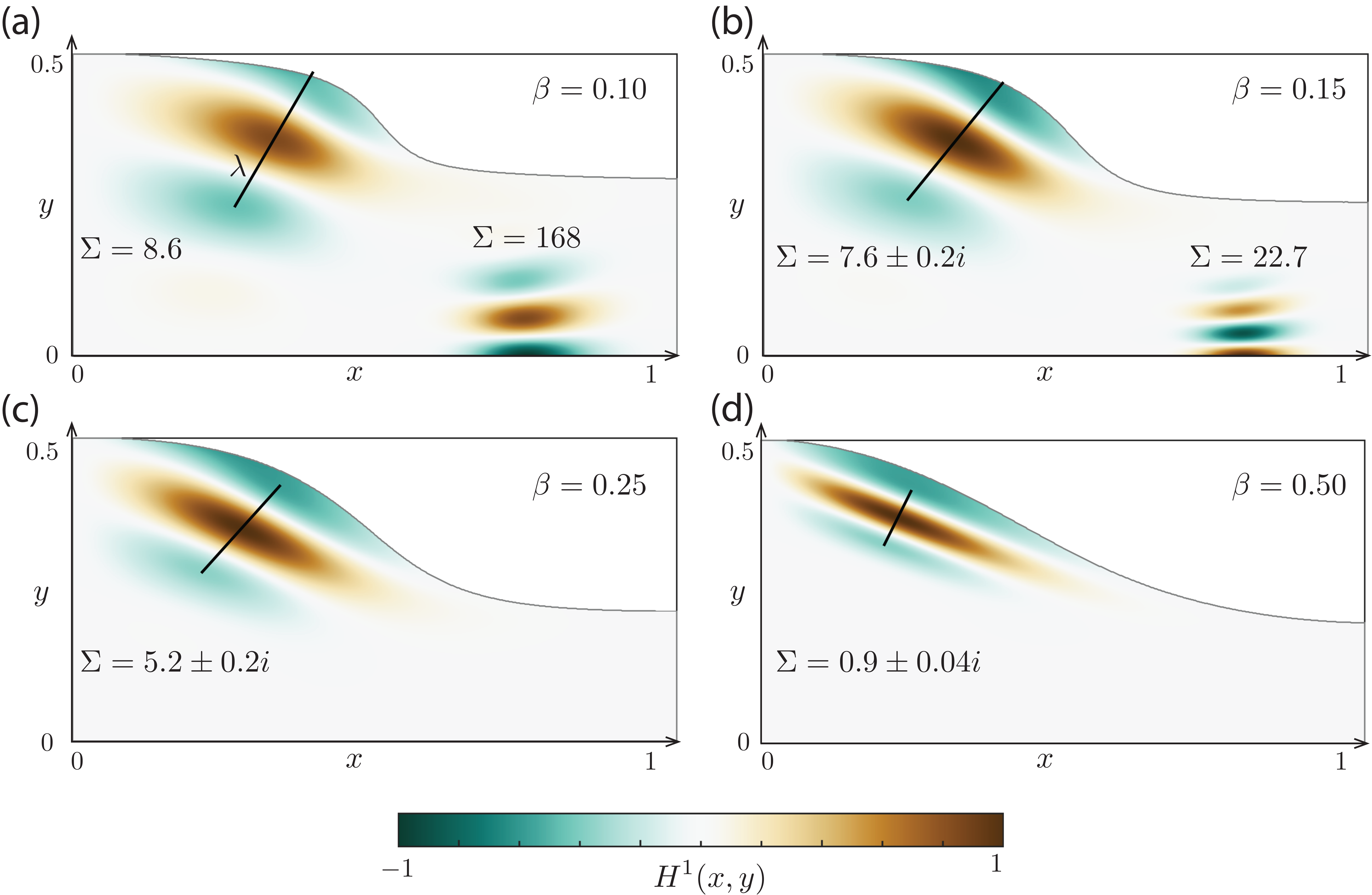}
\caption{Eigenmodes of the mid-plane deformation that are obtained by solving equation~(\ref{eq:eigenvalue_problem}) for values of (a) $\beta$=0.10 (b) $\beta$=0.15 (c) $\beta$=0.20 (d) $\beta$=0.50 and for a fixed stretching velocity $\alpha=8$ and $\Gamma=0$. Note that for each parameter value, the most unstable eigenmodes localized to the upstream and downstream region are combined and plotted in a single figure, as discussed in the main text. For $\beta>0.2$ there are no unstable modes downstream. The growth rates $\Sigma$ in the upstream region for the plots in (a)-(d) are $8.6$, $7.6 \pm 0.2 i$, $5.2 \pm 0.2i$ and $0.9 \pm 0.04i$ respectively. In the downstream region, $\Sigma=168.0$ for $\beta=0.10$ and $\Sigma=22.7$ for $\beta=0.15$.
}
\label{fig:out_of_plane}
\end{figure}

\subsection{Linear Stability}
\label{sec:linear stability}
In the previous section, we have calculated the base state stress distributions for the stretching flow of the viscous sheet where the mid-surface of the sheet $H(x,y)=0$ is assumed to be planar and constant. The existence of zones with compressive stresses implies that the constant mid-surface solution can be unstable to out-of-plane deformations in the absence of any stabilizing forces. We linearize (\ref{eq:BNT_dimless_a})-(\ref{eq:BNT_dimless_c}) to derive the eigenvalue problem that governs the out-of-plane displacement. We consider small perturbations to the mid-plane of the form,

\begin{equation}
\label{eq:perturb_form}
H(x,y)=H^0+H^1(x,y)\exp(\Sigma t)
\end{equation}

where $H^0=0$ is the base state, $H^1(x,y)$ is the shape of the mode and $\Sigma$ is the growth rate, with $\text{Re}(\Sigma)>0$ indicating unstable modes. Substituting this expression in (\ref{eq:T11})-(\ref{eq:M22}) and retaining terms that are linear in $H^1(x,y)$ leads to $T_{11}=2 \mu h(2u_x +v_y)$, $T_{12}=\mu h(u_y +v_x)$ and $T_{22}=2 \mu h(u_x +2v_y)$, i.e. upon linearization, the in-plane z-integrated stresses that arise during the incipient out-of-plane deformations of the thin sheet are identical to the base state stresses during stretching flow, to first order in $H^1$. Linearizing the out-of-plane force balance of (\ref{eq:BNT_dimless_c}) that balances the transverse shear stress and bending moments and retaining terms that are first order in $H^1$ leads to the eigenvalue problem for the out-of-plane deformations,

\begin{multline}
\label{eq:eigenvalue_problem}
\Sigma\Biggl\{\biggl[\Bigl(\frac{\mu h^3}{6}\Bigr)(2H^1_{xx}+H^{1}_{yy} )\biggr]_{xx}+\biggl[\Bigl(\frac{\mu h^3}{6}\Bigr)(H^1_{xx}+2H^{1}_{yy}) \biggr]_{yy} +\biggl[\Bigl(\frac{\mu h^3}{3}\Bigr)H^1_{xy}\biggr]_{xy}\Biggr\}=\\
\begin{aligned}
&T_{11}H^{1}_{xx}+T_{22}H^{1}_{yy}+2T_{12}H^{1}_{xy}- \Gamma(H^1_{xx}+H^1_{yy})
\end{aligned}
\end{multline}

where $h(x,y)$, $T_{11}$, $T_{12}$ and $T_{22}$ are already known from the base state solution obtained by solving (\ref{eq:trouton_dimless_a})-(\ref{eq:trouton_dimless_b}) for a given value of $\alpha$ and $\beta$, and $\mu(x,y)$ is given by the prescribed temperature field in (\ref{eq:viscosity_field}). The linearized form of the boundary conditions in (\ref{eq:eigenprob_bc_a}) for the inlet and outlet at $x=0$ and $x=1$ are,
 
\begin{equation}
\label{eq:linearized_clamped_bc}
H^1=0 \quad \text{and} \quad H^1_x=0.
\end{equation}

At $y=0$, the symmetry boundary condition of (\ref{eq:eigenprob_bc_c}) reduces to,
\begin{equation}
\label{eq:symmetry_bc}
H^1_y=0 \quad \text{and} \quad T_{12}H^1_x+T_{22} H^1_y + \Gamma H^1_y -\Sigma \left\{\left[\left(\frac{\mu h^3}{3}\right) H^1_{xy}\right]_x+\left[\left(\frac{\mu h^3}{6}\right) (H^1_{xx}+2H^1_{yy})\right]_y\right\}=0.
\end{equation}

and the linearized free edge boundary conditions at the lateral edge $y=D(x)$ are,

\begin{equation}
(2H^1_{xx}+H^1_{yy})\cos^2\phi+2H^1_{xy}\cos\phi \sin\phi+(H^1_{xx}+2H^1_{yy})\sin^2\phi=0,
\end{equation}

\begin{equation}
\begin{split}
\label{eq:free_edge_bc}
\left[T_{11}H^1_x+T_{12}H^1_y+\Gamma H^1_x +\Sigma (M_{11})_x+ \Sigma (M_{12})_y\right]\cos\phi + \left[T_{12}H^1_x+T_{22}H^1_y+\Gamma H^1_y \right. \\
\left. + \Sigma (M_{12})_x+\Sigma (M_{22})_y\right]\sin\phi + \Sigma \left[(M_{22})_x-(M_{11})_x\right]\sin^2\phi \cos\phi+ \Sigma \left[(M_{11})_y-(M_{22})_y \right]=0.
\end{split}
\end{equation}

The most unstable eigenmode corresponds to the solution $H^{1}(x,y)$ that has largest value of the growth rate $\text{Re}(\Sigma)$. The eigenvalue $\Sigma$ appears both in (\ref{eq:eigenvalue_problem}) and in the boundary conditions via (\ref{eq:symmetry_bc}) and (\ref{eq:free_edge_bc}). We use the Comsol finite element package to solve for the the largest eigenvalue $\Sigma$ that corresponds to the most unstable eigenmode and the shape of the out-of-plane displacement $H^1(x,y)$ for that eigenmode by a implementing a cubic lagrange shape function discretization.

\subsubsection{The case when the inverse Capillary Number $\Gamma=2\hat{\gamma}/(\hat{\mu}_0 \hat{U}_0)=0$}

To determine the growth rates $\Sigma$ and dimensionless wavelength $\lambda$, we use the ansatz in (\ref{eq:perturb_form}) for the entire fluid sheet to solve the global eigenvalue problem defined in (\ref{eq:eigenvalue_problem}). The eigenmodes we obtain fall into two categories: (i) those with out-of-plane deformations localized only near the upper compressive zones that we refer to as `upstream modes', and (ii) those with deformations localized in the downstream compressive region that we refer to as `downstream modes'.  Each eigenmode that is localized in the upstream and downstream regions has a distinct growth rate. In Fig.~\ref{fig:out_of_plane}, we show the combined shape of the most unstable buckling eigenmodes $H^1(x,y)$ from each region for different values of $\beta = \hat{\beta}/\hat{L}$, and fixed $\alpha=\hat{U}_1/\hat{U}_0$, and with the inverse capillary number $\Gamma=0$. That is, we combine the fastest growing out-of-plane modes separately from the upstream and downstream solutions (\textit{i.e.}, with each having a distinct value of $\Sigma$) in a single figure, and plot the sum, with the maximum amplitude in each region normalized to unity.

\begin{figure}[t]
\centering
\subfloat[][Upstream Growth Rate]{\includegraphics[height=3.5cm]{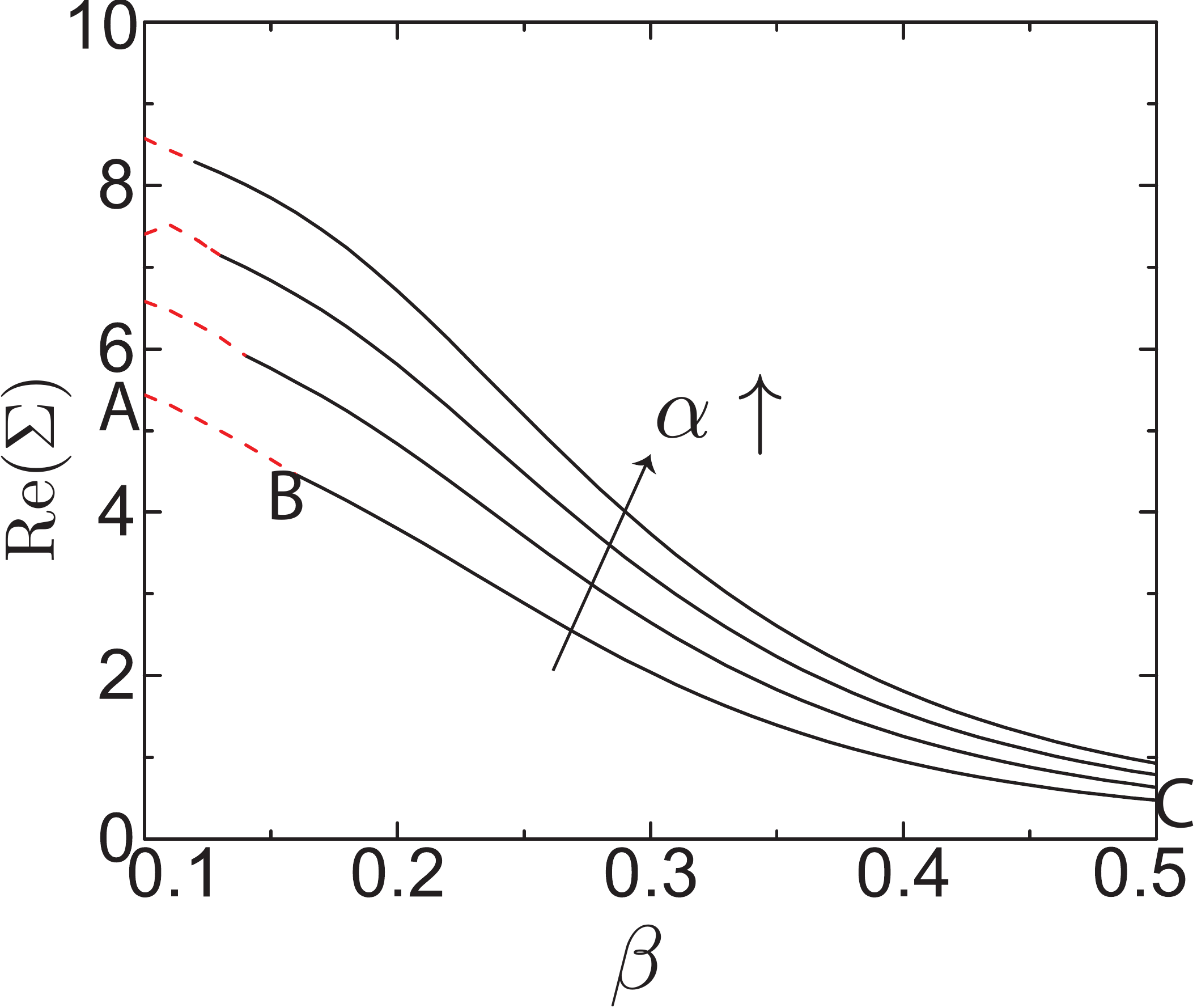}
\label{phase_map:a}}
\subfloat[][Upstream Phase Map]{\includegraphics[height=3.5cm]{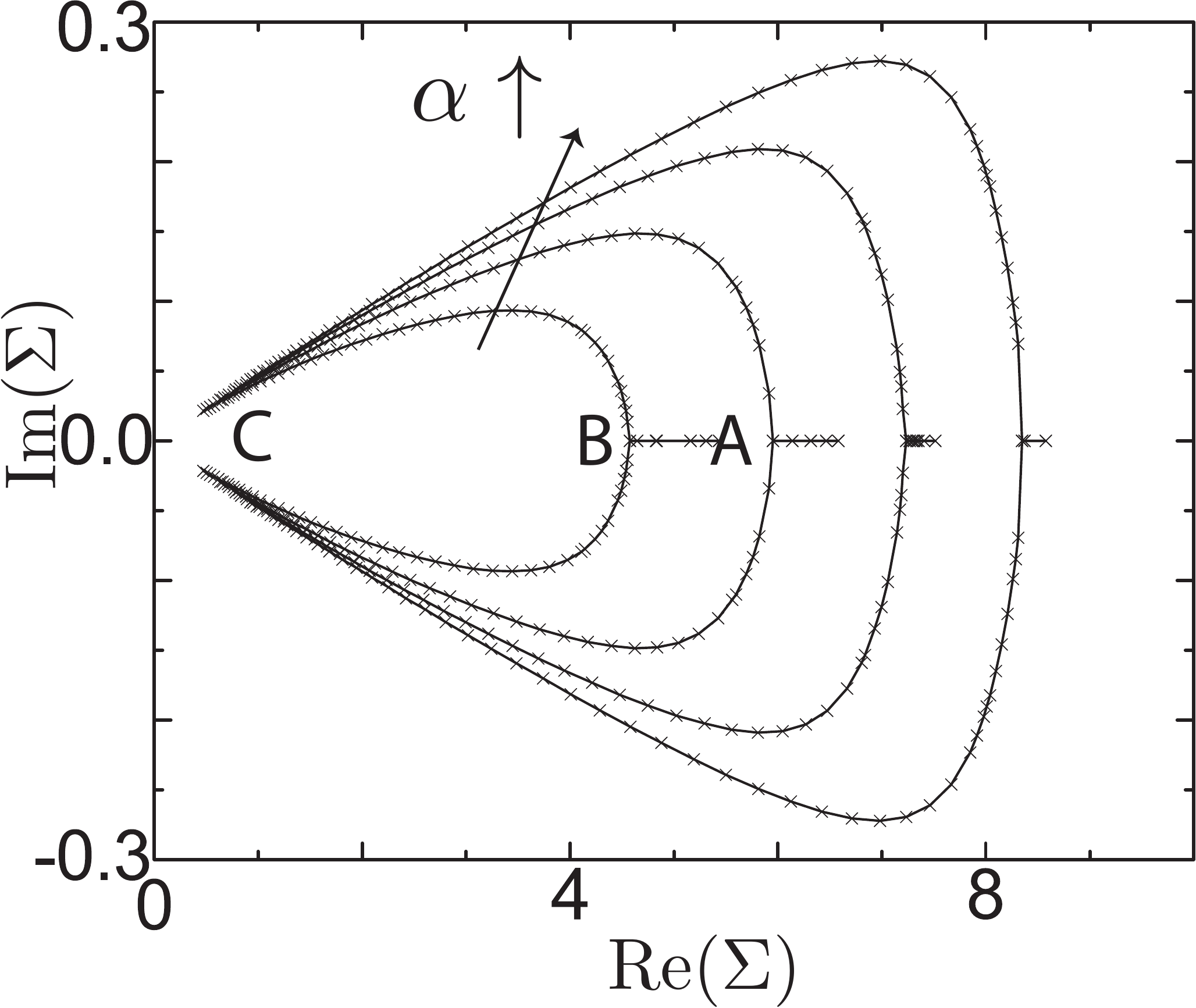}
\label{phase_map:b}}
\subfloat[][Downstream Growth Rate]{\includegraphics[height=3.5cm]{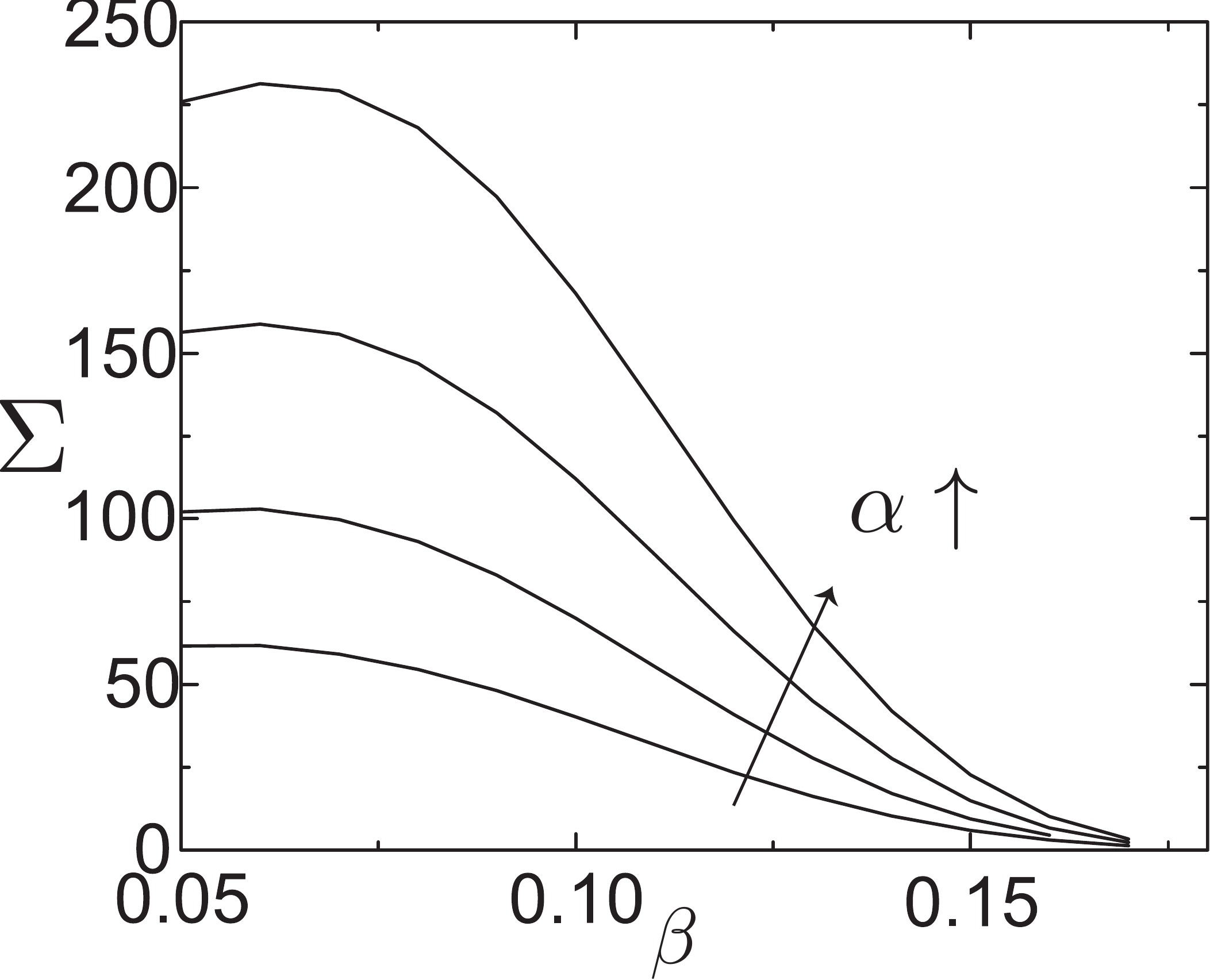}
\label{phase_map:c}}
\subfloat[][Upstream Wavelengths]{\includegraphics[height=3.5cm]{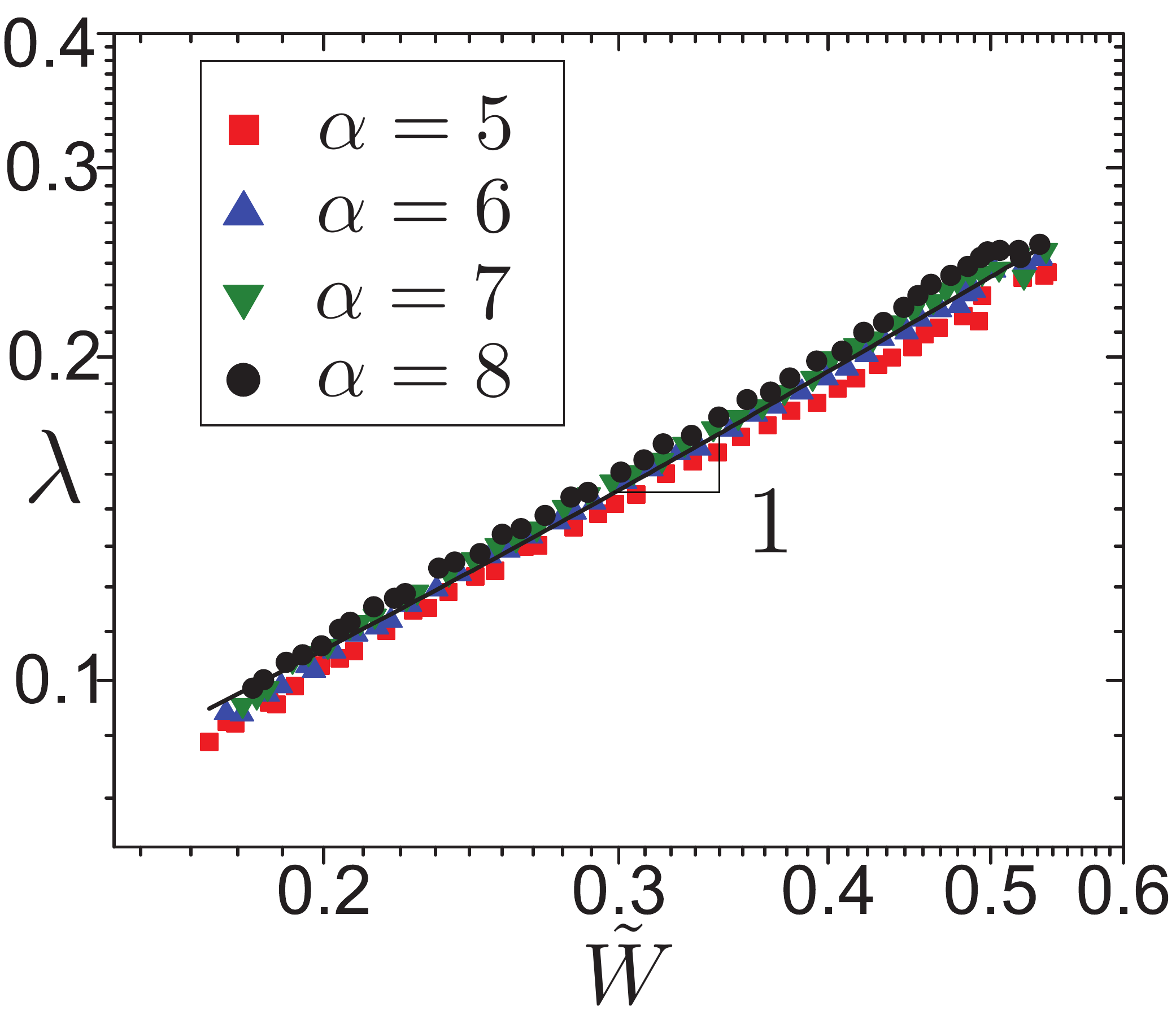}
\label{fig:wavelength}}
\caption{ (a) Real part of the growth rate $\Sigma$ of the most unstable upstream buckling mode in equation~(\ref{eq:perturb_form}) as a function of $\beta$. The red dashed lines are regions where $\Sigma$ is purely real and corresponds to stationary modes, while the solid black lines indicate growth rates with non-zero complex parts corresponding to oscillatory modes. (b) Corresponding phase diagram of the growth rate of the most unstable upstream eigenmode in the complex plane. The crosses represent discrete points where $\Sigma$ was determined for a pair of values $(\alpha,\beta)$. (c) Growth rate of the most unstable downstream buckling mode as a function of $\beta$. (d) The dimensionless wavelength $\lambda=\hat{\lambda}/\hat{L}$ (as shown in Fig.~\ref{fig:out_of_plane}) of the most unstable upstream mode with $\Gamma=0$ plotted against the effective width of the compression zone $\bar{W}$ for various values of $\alpha$ and $\beta$.}
\label{fig:phase_map}
\end{figure}

In the absence of surface tension, the upstream zone of the sheet is always unstable and undergoes buckling at all shear rates. The location of these eigenmodes overlaps well with the regions of compressive stress in the base state shown in Fig.~\ref{fig:stresses}, as expected. A phase map of the growth rates of the most unstable mode in the upstream region can be determined as shown in Fig.~\subref*{phase_map:a} and \subref*{phase_map:b}, where $\text{Re}(\Sigma)$ decreases with increasing $\beta$ and increases with increasing $\alpha$. In general, the non-hermitian form of (\ref{eq:eigenvalue_problem}) allows for oscillatory modes where $\text{Im}(\Sigma)\neq 0$. \\

To show the transition between stationary and oscillatory modes in the upstream region, we consider a path starting at the point A in Fig.~\subref*{phase_map:a}. For small values of $\beta$ (\textit{i.e.}, in the region in Fig.~\subref*{phase_map:a} denoted by dashed lines), we observe that $\text{Re}(\Sigma)>0$ and $\text{Im}(\Sigma)=0$, leading to stationary instabilities that span the upstream region of the viscous sheet. An example of a stationary unstable mode is shown in Fig.~\ref{fig:out_of_plane}(a). Gradually increasing $\beta$ (\textit{i.e.}, following the dotted red line in Fig.~\subref*{phase_map:a}) continues to result in stationary instabilities until the bifurcation threshold is reached at point B. Beyond this critical point, we observe that $\text{Re}(\Sigma)>0$ and $\text{Im}(\Sigma)\neq 0$ resulting in a pair of complex conjugate eigenvalues that result in oscillatory unstable modes. This change in behavior corresponds to a transition from a stationary bifurcation to a Hopf bifurcation. By (\ref{eq:perturb_form}), these complex valued growth rates correspond to a pair of unstable traveling modes with a period $T=2\pi/\text{Im}(\Sigma)$. Example deformation shapes of these oscillatory modes are shown in Fig.~\ref{fig:out_of_plane}(b)-(d). This unstable oscillatory regime is in contrast to the case of a homogeneous elastic sheet \cite{Southwell582}, where only stationary modes are allowed. The transition to the unstable oscillatory mode occurs at earlier values of $\beta$ for increasingly large stretching velocities $\alpha$, as shown in Fig.~\subref*{phase_map:b}. In contrast, the downstream unstable eigenmodes are always purely real, and therefore stationary for all values of $\alpha$ and $\beta$ in our study. The growth rates downstream are also seen to be larger in magnitude than the upstream region. Additionally, the growth rates downstream sharply drop with increasing $\beta$ until they reach 0 as seen in Fig.~\subref*{phase_map:c}, after which the downstream compressive zone entirely vanishes. 

The shape of the eigenmodes we obtain, and the transition from a static to a Hopf bifurcation, is analogous to the curtain modes seen in the work of Perdigou \& Audoly \cite{Perdigou2016291} for a falling viscous sheet. Similar to their results, we observe that the wavelength of the instability does not span the entire width of the sheet, and is instead localized in the compressive zones. In the upstream region, the wavevector is perpendicular to the edge of the sheet. The wavelength also sharply decreases upon increasing the width of the heating zone. This phenomena, seen in Fig.~\ref{fig:out_of_plane}(a)-(d), is directly related to the reduction in size of the local compressive region in the base state shown in Fig.~\ref{fig:stresses}(a)-(d). We define $\tilde{W}$ as the length of the longest chord that lies within the compressive zone, passing through the location of maximum compressive stress and oriented along the direction of compression. To better illustrate this dependence of the wavelength on the size of the compressive zone, we plot the spatial wavelength of the most unstable mode in Fig.~\subref*{fig:wavelength}, against the effective width $\tilde{W}$ of the compressive zone along the wavevector direction. $\lambda$ is observed to be linearly proportional to $\tilde{W}$ for all values of draw ratio, confirming that the deformation in the upstream region occurs by a buckling instability. In the downstream region, the unstable eigenmodes vanish beyond a value of $\beta\approx 0.2$, even when $\Gamma=0$, when the viscous sheet locally transitions from a state of being under local compression to under local tension as discussed earlier.

\subsubsection{The case when the inverse Capillary Number $\Gamma\neq 0$}

\begin{figure}[!htb]
\centering
\includegraphics[height=6cm]{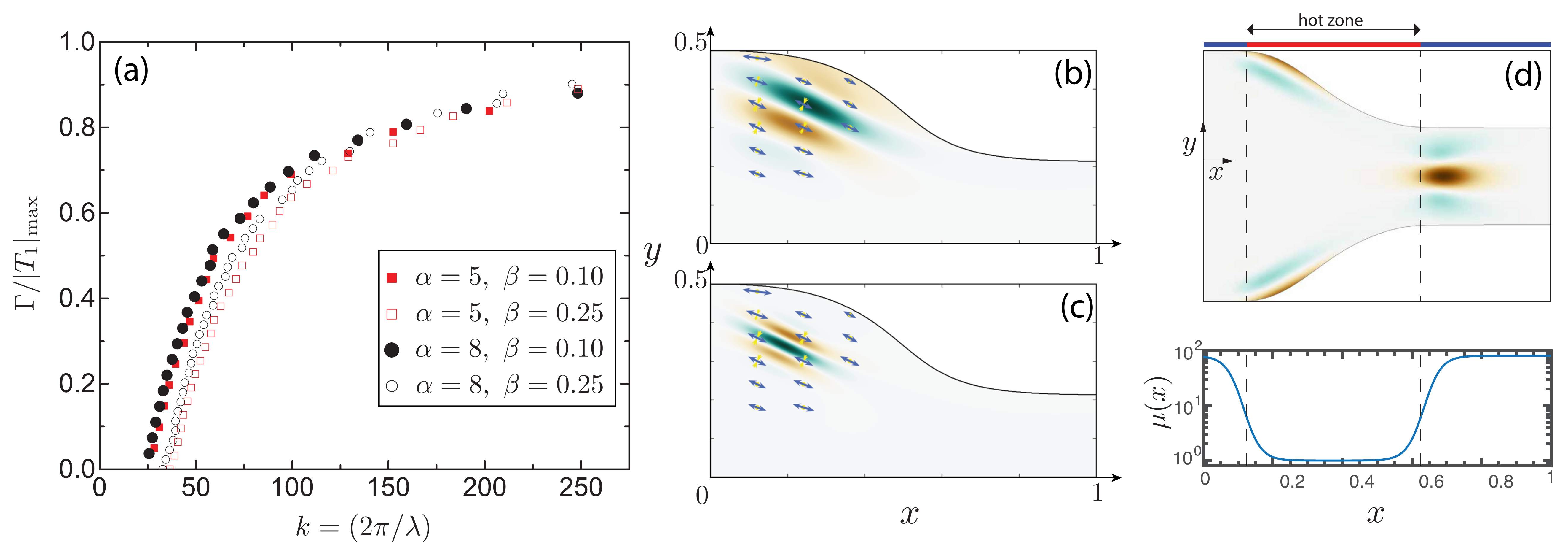}
\caption{\label{fig:surface_tension} (a) Variation of the most unstable wavenumber $k=2\pi/\lambda$ in the upstream zone upon increasing the inverse capillary number $\Gamma=\hat{\gamma}/(\hat{\mu}_0 \hat{U}_0)$ scaled by the maximum magnitude of the compressive stress $|T_1|_\text{max}$. (b,c) Most unstable modes for $\Gamma=1$ and $\Gamma=3$ respectively, with $\alpha=8$ and $\beta=0.25$. (d) Most unstable eigenmodes for operating conditions used in a prototypical industrial redraw process, where $\alpha=10$ and $\Gamma=0.0015$. The viscosity field $\mu(x)$ is plotted according to (\ref{eq:temp_viscosity_ind}), with parameter values $\theta_0=0.8$, $x_L=0.125$, $x_R=0.625$, $k=40$, $T_c=0.34$ and $\nu=0.15$.}
\end{figure}

If the compressive membrane stresses exceeds the inverse capillary number, all wavelengths will be unstable, and surface tension will not selectively stabilize the small wavelength deformations. This happens when the rate at which the surface tension flattens the deformation of the midsurface is less than the rate at which compressive stresses destabilize the sheet. In the viscous plate eigenvalue problem (\ref{eq:eigenvalue_problem}), the stabilizing effect of surface tension (via $\Gamma$) and the destabilizing effect of the compressive membrane stresses, are both coupled to the curvature of the mid-surface and scale as $\lambda^{-2}$. Therefore, consistent with this physical interpretation, we observe that the viscous sheet is linearly stable to all out-of-plane deformations in the limit when $\Gamma>|T_1|_\text{max}$. In Fig.~\ref{fig:surface_tension}(a), the most unstable wavenumber $k=2\pi/\lambda$ is plotted for various values of $\Gamma/|T_1|_\text{max}$. The curves are generated by varying $\Gamma$ for the specified value of $\alpha$ and $\beta$ indicated in the legend, and estimating $\lambda$ and $|T_1|_\text{max}$ from the solutions of the corresponding most unstable mode and the base state as discussed in the previous sections. For each of these curves, we see that $k\rightarrow \infty$ as $\Gamma$ approaches $|T_1|_\text{max}$.

As the long wavelength deformations are only weakly aligned to the compressive membrane stress eigenvector direction, a finite value of $\Gamma$ is sufficient to suppress these long wavelength deformations and instead select for short wavelength modes that are strongly aligned to the compressive eigenvector direction. This effect can be seen in Fig.~\ref{fig:surface_tension}(b) and (c), where increasing $\Gamma$ leads to smaller wavelengths whose wavevectors are aligned with the compressive stress eigenvector indicated by the yellow arrows. 

\subsubsection{Instability of a glass sheet under redraw}

To provide an example of how this numerical solution method can be applied to the glass redraw process, we simulate the base state flow and determine the stability of the viscous sheet in an industrially relevant scenario. Typical furnace heating profiles that are currently used exhibit sharp gradients in the fluid sheet temperature near the heating zone, such that the sheet temperature rapidly varies from its inlet value to a maximum value over a narrow region. To mimic these real operating conditions, and to show the applicability of our technique beyond the simple Gaussian model used in (\ref{eq:temp_profile}), we use the temperature-viscosity relation implemented in \cite{kiely_jfm_2015},

\begin{equation}
\begin{gathered}
\label{eq:temp_viscosity_ind}
T(x)=\theta_0+(1-\theta_0)\left(\frac{1}{1+\exp\left[-k(x-x_L)\right]}+\frac{1}{1+\exp\left[k(x-x_R)\right]}-1\right), \\
\mu(x)=\exp\left[\frac{1}{\nu}\left(\frac{1}{T-T_c}-\frac{1}{1-T_c} \right) \right].
\end{gathered}
\end{equation}

Here, $\theta_0=0.8$, $x_L=0.125$, $x_R=0.625$, $k=40$, $T_c=0.34$ and $\nu=0.15$ are the relevant parameters that result in a dimensionless viscosity profile that is shown in Fig.~\ref{fig:surface_tension}(d), where $x=\hat{x}/\hat{L}$. From Fig.~\ref{fig:schematic}(a), we estimate the ratio of the length to the width of the sheet to be $\hat{W}/\hat{L}=0.72$. We assume a draw ratio of $\alpha=10$, and a surface tension value of $\gamma \approx 250$ mN/m, corresponding to an inverse capillary number of $\Gamma=0.0015$. The small value of $\Gamma$ confirms that the large viscous stresses dominate capillary effects in the glass redraw process.  \\

Under these conditions, the viscous glass sheet is unstable to out-of-plane deformations, and the most unstable eigenmodes are shown in Fig.~\ref{fig:surface_tension}(d), with the dimensionless growth rate $\Sigma=2.2$ in the upstream region and $\Sigma = 83$ in the downstream region. Comparing the results of our simulation in Fig.~\ref{fig:surface_tension}(d) with Fig.~\ref{fig:schematic}(a) demonstrates that we are able to qualitatively reproduce the shape of the viscous sheet and the location of the buckling eigenmodes in both the upstream and downstream regions. Our results are sensitive to the prescribed temperature field and therefore requires precise determination of the temperature profile and viscosity in the sheet in order to quantitatively compare the sheet thickness, wavelength and growth rates. As our model only considers the initial linear out-of-plane deformation, we cannot predict the non-linear evolution of the instability and its final shape and amplitude. Despite these limitations, the results of our simulation are consistent with the final shape and deformation patterns seen in Fig.~\ref{fig:schematic}(a), and demonstrate the potential use in the redraw process, particularly in the context of avoiding wrinkles, a problem we now turn to.

\subsection{Inverse problem: how to prevent wrinkles}
\label{sec:inverse_problem}
\begin{figure}[!htb]
\centering
\includegraphics[width=\columnwidth]{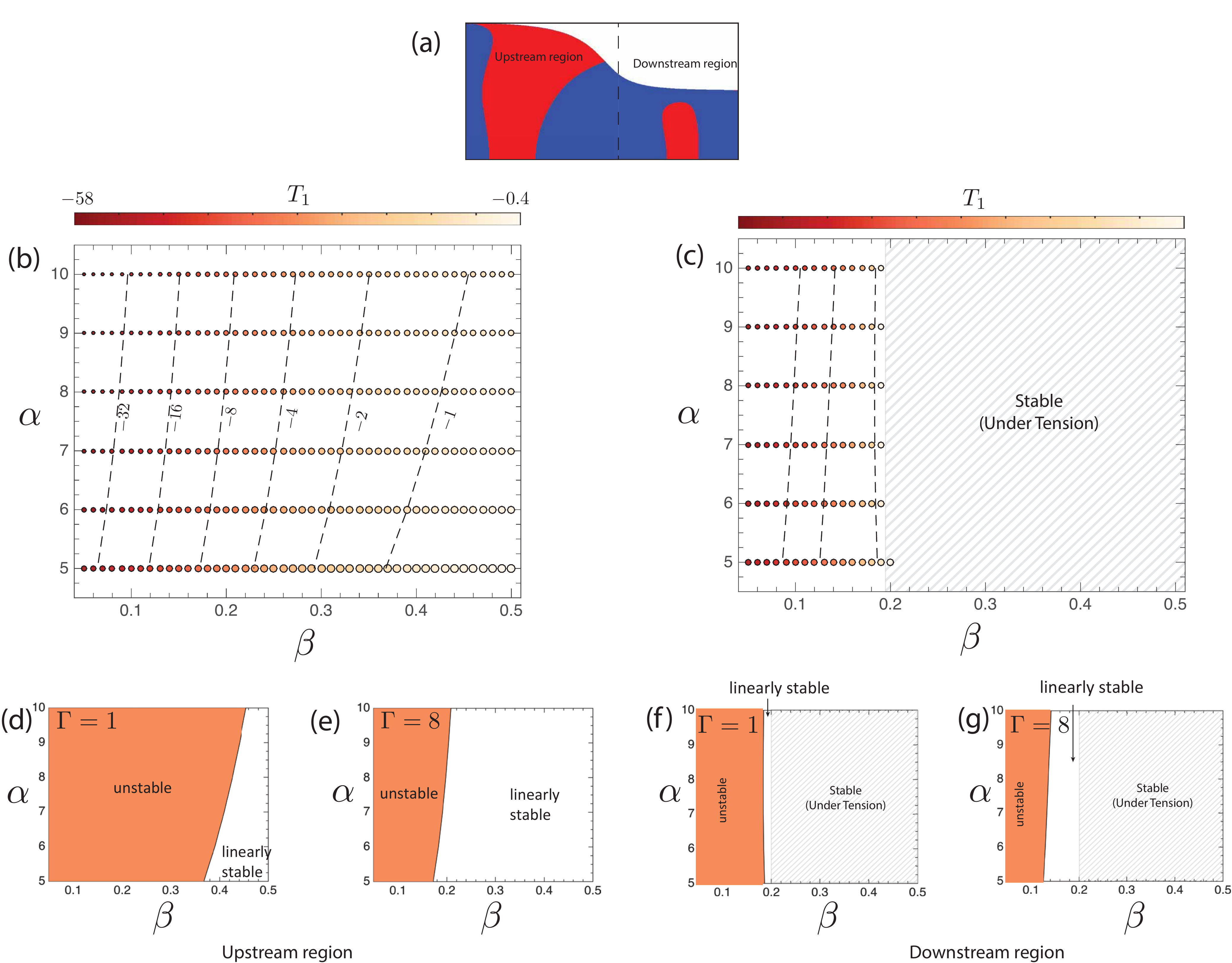}
\caption{\label{fig:inverse_combined} (a) Location of the zones of compressive stress in the upstream and downstream regions obtained by solving (\ref{eq:continuity})-(\ref{eq:trouton_dimless_b}) for the base state. Regions with compressive stresses are shown in red, and regions under tension in blue. (b,c) Operating diagram where each circle corresponds to a base-state solution of (\ref{eq:continuity})-(\ref{eq:trouton_dimless_b}) for a fixed $(\alpha,\beta)$ where $5\leq \alpha \leq 10$ in steps of unity and $0.05 \leq \beta \leq 0.5$ in steps of 0.01 for the (b) upstream and (c) downstream regions respectively. The size of each circle is proportional to the mean sheet thickness $\langle h \rangle$ at $x=1$ and the color denotes $T_1$, the magnitude of the minimum in-plane stress. The dotted lines correspond to contour lines of fixed values of $T_1$. The hatched region correspond to parameter values where the downstream zone is under tension (and thus globally stable) for all values of $\alpha$ and $\Gamma$. (d,e) Stability diagrams for (d) $\Gamma=1$ and (e) $\Gamma=8$ respectively in the upstream region. (f,g) Stability diagrams for (f) $\Gamma=1$ and (g) $\Gamma=8$ respectively in the downstream region.}
\end{figure}
By eliminating the downstream compressive zone, the out-of-plane deformations of the glass sheet near the outlet can be suppressed. This leads to the inverse problem, \textit{i.e.} of determining the set of values in the parameter space of draw ratios $\alpha$ and choice of furnace heating profile, that either eliminates the buckling instability or minimizes the out-of-plane deformation, while simultaneously achieving a target sheet thickness $\langle h \rangle$ at the outlet. In order to parameterize the temperature profile in a simple way, we use the Gaussian temperature profile defined in (\ref{eq:temp_profile}) and vary $\beta$, the width of the heating zone. Identifying regimes in parameter space of $\alpha, \beta, \Gamma$ that eliminate out-of-plane instabilities, if they exist, would allow for a rational framework to manufacture ultrathin sheets of glass by the redraw method. 
  
In Fig.~\ref{fig:inverse_combined}(b) and (c), we provide a design chart for operating the redraw method, that corresponding to the upstream and downstream region respectively. Each circle corresponds to solving the base state equations (\ref{eq:continuity})-(\ref{eq:trouton_dimless_b}) for a fixed value of $\alpha$ and $\beta$, with the size of the circle proportional to the mean sheet thickness. To achieve the thinnest sheets at the outlet, large draw ratios $\alpha \gg 1$ and narrow heating zones $\beta \ll 1$ are required. This corresponds to the upper left region of Fig.~\ref{fig:inverse_combined}(b) and (c). However, a choice of $\alpha \gg 1$ and $\beta \ll 1$ results in a large values of that base state compressive stress $T_1$ that destabilizies the viscous sheet. The results of Section \ref{sec:linear stability} imply that the sheet is linearly stable to out-of-plane deformations when either $T_1>0$ or when $\Gamma>|T_1|_\text{max}$. When there is no surface tension, then all compressive regions undergo a wrinkling instability. Increasing the magnitude of $\Gamma$ can stabilize regions of the sheet that were previously unstable to out-of-plane deformations. However, for the glass redraw process, the viscous forces are generally much larger than the capillary forces, and consequently $\Gamma \ll 1$ and it is unrealistic to achieve $\Gamma \sim |T_1| \sim O(1)$ using the stretching velocities normally employed. Therefore, in practical applications, the destabilizing effect of the viscous stresses cannot be stabilized by surface tension alone. We show the influence of surface tension for completeness in Fig.~\ref{fig:inverse_combined}(d)-(g). The first two plots show regions of the parameter space in the upstream region that are linearly stable due to surface tension, where $\Gamma>|T_1|_\text{max}$. The last two plots show a similar transition to stability for the downstream region with increasing $\Gamma$. Note that $\beta>0.2$ is globally stable even when $\Gamma=0$ as the sheet is under tension, and this transition to stability appears to be insensitive to the choice of $\alpha$. The shaded region in Fig.~\ref{fig:inverse_combined}(b), highlights the region of parameter space where the sheet is always under tension in the downstream region.
 
Our analysis therefore immediately suggests a method to manufacture ultra-thin glass sheets of a required thickness. As the stability boundary in Fig.~(\ref{fig:inverse_combined}) can only be slightly shifted by manipulating the surface tension of the molten glass, the optimal strategy in the redraw method is to entirely eliminate the downstream regions of compressive stress by utilizing a sufficiently wide, gradually varying heating profile (\text{e.g.} choosing $\beta>0.2$ in Eq.~(\ref{eq:temp_profile})), coupled with large values of draw ratio $\alpha>10$ to achieve the desired target output thickness.

\section{Conclusions}

The manufacturing of glass for electronics applications requires the processing of thin viscous sheets which are susceptible to wrinkling instabilities. Here, we have analyzed this instability by considering the deformation of the midsurface of thin viscous sheets with a non-homogeneous temperature field  for a sheet thickness and viscosity subject to an extensional flow within the framework of a linearized thin viscous plate model (\textit{c.f.}, (\ref{eq:eigenvalue_problem})). The effect of the stretching velocity is characterized by two dimensionless parameters: an outlet/inlet velocity ratio parameter $\alpha$ and a scaled  width of the heating zone $\beta$. The extensional flow induced by stretching at the outlet, coupled with the localized heating zone around midpoint of the sheet, leads to a lateral contraction and reduction of the sheet thickness. Localized zones of compressive stresses develop in two regions that are respectively upstream and downstream of the heating zone, that tend to destabilize the sheet in the absence of surface tension.   In the upstream region, the instability was shown to manifest either as a stationary or an oscillatory mode, depending on the value of $\alpha$ and $\beta$. In contrast, the downstream unstable modes are always stationary with purely real growth rates. Additionally, the downstream unstable mode was shown to vanish beyond a critical width of $\beta \approx 0.2$, where the state of stress locally transitions from compression to tension. Including surface tension stabilizes the sheet when the inverse capillary number $\Gamma$ is larger than the maximum magnitude of the compressive stress. Finally, we develop an engineering diagram to aid in the choice of outlet draw velocity $\alpha$ and heating zone width $\beta$ that result in desired normalized sheet thickness while still maintaining stability. Our framework can be readily extended to include more complicated spatially inhomogeneous flows,  and therefore is beneficial in studying the dynamics of thin sheets in many diverse processes in physical and biological settings. 

\begin{acknowledgments}
We would like to acknowledge N. Kaplan for helpful comments, discussions and suggestions with the finite element COMSOL simulations. We thank M. Nishikawa and the Asahi Glass Company for sharing the experimental image (Fig.~\ref{fig:schematic}(a)) of the wrinkled glass sheet.
\end{acknowledgments}

\bibliography{references}
\end{document}